\documentclass{emulateapj}
\shorttitle{Untriggered GRB Afterglows}
\shortauthors{Rykoff et al.}

\begin{document}

\title{A Search for Untriggered GRB Afterglows with ROTSE-III}

\author{Rykoff,~E.~S.\altaffilmark{1}, 
Aharonian,~F.\altaffilmark{2},
Akerlof,~C.~W.\altaffilmark{1},
Alatalo,~K.\altaffilmark{1},
Ashley,~M.~C.~B.\altaffilmark{3}, 
G\"{u}ver, T.\altaffilmark{4},
Horns,~D.\altaffilmark{2},
Kehoe,~R.~L.\altaffilmark{5},
K{\i}z{\i}lo\v{g}lu,~\"{U}.\altaffilmark{6},
McKay,~T.~A.\altaffilmark{1},
\"{O}zel,~M.\altaffilmark{7},
Phillips,~A.\altaffilmark{3}, 
Quimby,~R.~M.\altaffilmark{8},
Schaefer,~B.~E.\altaffilmark{9}, 
Smith,~D.~A.\altaffilmark{1},
Swan,~H.~F.\altaffilmark{1},
Vestrand,~W.~T.\altaffilmark{10}, 
Wheeler,~J.~C.\altaffilmark{8},
Wren,~J.\altaffilmark{10}, 
Yost,~S.~A.\altaffilmark{1} }

\altaffiltext{1}{University of Michigan, 2477 Randall Laboratory, 450 Church
        St., Ann Arbor, MI, 48104, erykoff@umich.edu, akerlof@umich.edu,
        kalatalo@umich.edu, tamckay@umich.edu, donaldas@umich.edu,
        hswan@umich.edu, sayost@umich.edu}
\altaffiltext{2}{Max-Planck-Institut f\"{u}r Kernphysik, Saupfercheckweg 1,
        69117 Heidelberg, Germany, Felix.Aharonian@mpi-hd.mpg.de,
        horns@mpi-hd.mpg.de}
\altaffiltext{3}{School of Physics, Department of Astrophysics and Optics,
        University of New South Wales, Sydney, NSW 2052, Australia,
        mcba@phys.unsw.edu.au, a.phillips@unsw.edu.au}
\altaffiltext{4}{Istanbul University Science Faculty, Department of Astronomy
        and Space Sciences, 34119, University-Istanbul, Turkey, 
        tolga@istanbul.edu.tr}
\altaffiltext{5}{Dept. of Physics, Southern Methodist University, Dallas, TX
        75275, kehoe@physics.smu.edu}
\altaffiltext{6}{Middle East Technical University, 06531 Ankara, Turkey,
        umk@astroa.physics.metu.edu.tr}
\altaffiltext{7}{\c{C}anakkale Onsekiz Mart \"{U}niversitesi, Terzio\v{g}lu
        17020, \c{C}anakkale, Turkey, me\_ozel@ibu.edu.tr}
\altaffiltext{8}{Department of Astronomy, University of Texas, Austin, TX
        78712, quimby@astro.as.utexas.edu, wheel@astro.as.utexas.edu}
\altaffiltext{9}{Department of Physics and Astronomy, Louisiana State
        University, Baton Rouge, LA 70803, schaefer@lsu.edu}
\altaffiltext{10}{Los Alamos National Laboratory, NIS-2 MS D436, Los Alamos, NM
        87545, vestrand@lanl.gov, jwren@nis.lanl.gov}

\begin{abstract}
We present the results of a search for untriggered gamma-ray burst (GRB)
afterglows with the Robotic Optical Transient Search Experiment-III (ROTSE-III)
telescope array.  This search covers observations from September 2003 to March
2005.  We have an effective coverage of $1.74\,\mathrm{deg}^2\,\mathrm{yr}$ for
rapidly fading transients that remain brighter than $\sim17.5$ magnitude for
more than 30 minutes.  This search is the first large area survey to be able to
detect typical untriggered GRB afterglows.  Our background rate is very low and
purely astrophysical.  We have found 4 previously unknown cataclysmic variables
(CVs) and 1 new flare star.  We have not detected any candidate
afterglow events or other unidentified transients.  We can place an upper limit
on the rate of fading optical transients with quiescent counterparts dimmer
than $\sim\,20^{th}$ magnitude at a rate of less than
$1.9\,\mathrm{deg}^{-2}\,\mathrm{yr}^{-1}$ with 95\% confidence.  This places
limits on the optical characteristics of off-axis (orphan) GRB afterglows.  As
a byproduct of this search, we have an effective
$\sim52\,\mathrm{deg}^2\,\mathrm{yr}$ of coverage for very slowly decaying
transients, such as CVs.  This implies an overall rate of outbursts from high
galactic latitude CVs of $0.1\,\mathrm{deg}^{-2}\,\mathrm{yr}^{-1}$.
\end{abstract}

\section{Introduction}
\label{sec:intro}

There is much circumstantial evidence that gamma-ray burst (GRB) outflows are
highly relativistic and collimated.  An achromatic break has been seen in light
curves for several GRB afterglows, with the canonical example being
GRB~990510.~\citep{hbfsk99, sgkpt99} These breaks are naturally explained by a
geometric constraint on the outflow.  The jet opening angle has been inferred
for GRB~990510 and several other GRBs, and appears to range from $2^{\circ}$ to
$30^{\circ}$.~\citep{fksdb01, pk01} Therefore, the true rate of GRBs must be
$\gtrsim100$ times that detected by satellite experiments such as BATSE,
HETE-2, INTEGRAL, and Swift.  Although the Earth will not receive $\gamma$-ray
emission from these bursts, they might be detectable at longer wavelengths.  It
remains an open question what these off-axis ``orphan'' afterglows should look
like. In the simplest model, an orphan afterglow looks like a standard
afterglow, except that it becomes visible after a delay of
$\sim0.5\,\mathrm{day}$.  As the ejecta cools, the relativistic beaming angle
increases until the afterglow can be seen off-axis, long after the
$\gamma$-rays have ceased.  \citep{rhoad97} However, this assumes that there is
no significant optical emission outside the $\gamma$-ray beaming angle.
\citet{np03} have suggested that the beaming angle of the optical emission
might be different than that of the $\gamma$-ray emission.  They refer to these
afterglows as ``on-axis orphan afterglows.''  A detectable rate of orphan
afterglows gives an orthogonal approach to measuring typical GRB collimation.

Whether or not orphan (off-axis) afterglows are detectable, there must be
untriggered afterglows from normal GRBs that have simply not been seen by a
$\gamma$-ray satellite.  The Swift satellite can detect approximately two GRBs
per week in its field of view, while an extrapolation of the BATSE event
trigger rate for the entire sky suggests that there are around 2 GRBs per day
visible to the Earth, corresponding to
$0.018\,\mathrm{deg}^{-2}\,\mathrm{yr}^{-1}$.~\cite{fehkl93} How much solid
angle would a survey need to cover to observe an untriggered GRB afterglow
serendipitously?  We know from ROTSE-I and LOTIS prompt follow-ups to BATSE
triggers that the preponderance of early afterglows do not get as bright as
$14^{th}$ magnitude. \citep{abbbb00,kabbb01,ppwab99} Currently, we do not have
enough data on early afterglows with deeper imaging to provide any firm
predictions of the rate of detectable bursts.  We expect this to change now
that Swift is operational.  Rapid follow-up to a small number of HETE-2
triggers has shown that approximately 50\% of bursts might be brighter than
$\sim18.5$ for 30 minutes or more.~\citep{lrabb04} Thus, around
$9\times10^{-3}\,\mathrm{bursts}\,\mathrm{deg}^{-2}\,\mathrm{yr}^{-1}$ should
be visible to an instrument capable of reaching this magnitude, such as
ROTSE-III.  Thus, an exposure of $\sim110\,\mathrm{deg}^2\,\mathrm{yr}$ is
required for a high probability of finding an afterglow independent of any
$\gamma$-ray trigger.

To date, there have only been a few published searches for untriggered and
orphan GRBs and other short duration transients.  These searches have all
probed different magnitude ranges and timescales.  In general, the wide-field
instruments cover more solid angle but cannot go very deep.  The ROTSE-I
transient search covered $3.5\,\mathrm{deg}^2\,\mathrm{yr}$, to a limiting
magnitude of 15.7. \citep{kabbc02} The RAPTOR array covers the entire visible
sky several times each night to a limiting magnitude of 12, and is sensitive to
very fast transients, on the order of minutes. \citep{vbcfg04} The Deep Lens
Survey (DLS) transient search covered $0.01\,\mathrm{deg}^2\,\mathrm{yr}$ with
sensitivity to $24^{th}$ magnitude, and found a couple of tantalizing
unidentified transients. \citep{bwbcd04} In spite of the relatively rapid
detection of these transients, they were too faint for spectroscopic follow-up
and could not be positively identified as extragalactic or associated with
GRBs.  \citet{vlwbl02} performed a color-selected transient search with the
Sloan Digital Sky Survey to a limiting magnitude of 19 and detected one unusual
AGN.

The search reported in this paper was specifically designed to detect
untriggered and orphan GRB afterglows.  This search is based upon the
assumption that an orphan afterglow might have an optical behavior similar to
that of observed afterglows.  As a result, we search for transients which meet
two criteria: first, the quiescent counterpart or host galaxy would have $m_R >
20$, and would not be detectable by ROTSE-III; second, the transient must be
brighter than our limiting magnitude for at least 30 minutes.  Other known
astrophysical sources fall into this category, including: cataclysmic variables
(CVs) and novae in the galactic halo that burst by $>2$ magnitudes; faint flare
stars that brighten on short timescales by several magnitudes; and active
galactic nuclei (AGN), blazars, and quasars that display optically violent
variability (OVV), occasionally flaring by several magnitudes on very short
timescales.

The rapid identification of new transients is essential for a search of this
nature.  Only spectroscopic follow-up can positively identify an orphan
afterglow or a new type of astrophysical phenomenon. As ROTSE-III can identify
transients while they are still relatively bright, this enables follow-up with
telescopes with modest apertures.

\section{Observations \& Data Reduction}

The ROTSE-III systems are described in detail in \citet{akmrs03}.  The
ROTSE-III telescopes are installed at four sites around the globe:
Coonabarabran, Australia; Ft. Davis, Texas; Mt. Gamsberg, Namibia; and
Bakirlitepe, Turkey.  They have a wide ($1\fdg85 \times 1\fdg85$) field of view
imaged onto an E2V $2048\times2048$ back-illuminated thinned CCD, and operate
without filters.  The camera has a fast readout cycle of 6~s.  The limiting
magnitude for a typical 60~s exposure is around $m_{R}\sim18.5$, which is well
suited for study of GRB afterglows during the first hour or more.  The typical
FWHM of the stellar images is $<2.5$ pixels ($8\farcs1$).

In September 2003 we initiated analysis of our nightly sky patrol images for
rapid identification of fast transients.  The region patrolled includes $370\times
3.4\,\mathrm{deg}^2$ fields in the equatorial stripe with declination of
$|\delta|<2\fdg64$.  To avoid field crowding and galactic opacity, all the
fields are at high galactic latitude with $|b|>30^{\circ}$.  Because asteroids
were a significant background early in our search, after May 2004 we only
imaged fields with ecliptic latitudes of $|\beta|>10^{\circ}$. These fields
were chosen for two primary reasons.  First of all, they are visible to all
four ROTSE-III locations, two of which are in the northern hemisphere, and two
of which are in the southern hemisphere.  Furthermore, this region has public
Sloan Digital Sky Survey (SDSS) data with 5 color imaging to well below our
limiting magnitude. \citep{aaaaa05} This allows calibration of our fields to a
set of well-measured stars, as well as providing easy identification of flares
from objects such as CVs and quasars.

Our standard observing sequence includes a pair of closely spaced 60 s
exposures, followed after 30 minutes by a second pair of 60 s exposures.  We
define a \emph{set} of images as four consecutive images taken with this
interval.  The second image of each pair is offset by $\sim 10$ pixels to
reduce the impact of bad pixels.  During bright moon conditions (lunar
illumination $>70\%$) we reduce the exposure length to 20~s to prevent the
background sky from saturating the images.  The 30 minute interval was chosen
to enhance sensitivity to rapidly fading transients while still allowing a
large solid angle coverage.  

After each image is recorded, it is dark subtracted and flat fielded by an
automated pipeline.  Twilight flats are generated every night, and are updated
for the pipeline on a monthly basis.  Our back illuminated thinned CCD imager,
combined with broadband filterless optics, consequently imposes an interference
fringe pattern on all images.  We therefore must correct for these fringing
effects.  The pattern is stable, although the amplitude varies with the
brightness of night sky lines.  We have created a fringe map for each CCD by
comparing a twilight flat image, which does not display a fringe pattern, with
a sky flat image, which does display a fringe pattern.  After flat fielding,
the sky pixels in the central subregion of the image are fit via linear
regression with the corresponding pixels in the fringe pattern.  The fringe
pattern is then scaled accordingly and subtracted from the image.  If the fit
is poor (reduced $\chi^2>3$), due to scattered moonlight or clouds in the
image, the fringe pattern is not subtracted.  In the worst case, the fringe
pattern can introduce photometric errors as large as $5\%$, and can produce
occasional false detections of faint objects.

The pipeline then runs SExtractor~\citep{ba96} to perform initial object
detection, measure centroid positions, and determine aperture magnitudes.  A
separate pipeline program written in IDL ({\tt idlpacman}) correlates the
object list with stars brighter than $15^{th}$ magnitude in the USNO~A2.0
catalog to determine an astrometric solution as well as an approximate
magnitude zero-point for the field.

The limiting magnitude of each image is estimated from the background noise,
FWHM of the point spread function (PSF), and the zero-point offset, which is
essentially a measure of the transparency of the sky.  With these three
values, we can estimate the magnitude at which we can detect a star 90\% of the
time with our SExtractor cuts in an uncrowded region of sky. As will
be shown below in \S~\ref{sec:analysis}, our detection efficiency declines
very rapidly in crowded regions.

After each pair of images is calibrated, we pair-match the object lists.
Objects that are detected in both of the pair of images are considered real.
All objects that are detected only in a single image are rejected.  This
strategy removes most spurious detections caused by cosmic rays, pixel defects,
satellite glints, and noise spikes.  However, some backgrounds still remain:
noise spikes usually due to imperfect fringe subtraction when the sky is not
fully transparent; some cosmic ray coincidences; and asteroids.

One of the great difficulties in calibrating a large area survey of this type
is the lack of standard stars uniformly distributed across the sky.  The
USNO~A2.0 catalog provides excellent astrometric solutions for any field as
there are typically $>1000$ stars with $R<15$.  However, the
photometric zero-points as determined from USNO A2.0 $R$-band magnitudes have
typical systematic errors of up to $0.30$ magnitude \citep{mcdgh98} making
field-to-field comparisons difficult.  

For around 90\% of our fields, we have significant overlap with SDSS 5-color
data.  We have thus decided to recalibrate all of our fields relative to the
SDSS $r'$-band, as the field-to-field variations are around
2\%. \citep{aaaaa05} For each of our overlapping fields we compare all
ROTSE-III template stars that have counterparts in SDSS between $15<r'<17$ with
$g'-r' < 1.0$.  We find that the typical offset is $0.22\pm0.16$ when
converting from $m_{\mathrm{ROTSE(USNO)}}$ to $m_{\mathrm{ROTSE(SDSS)}}$.  The
scatter in offsets is primarily due to systematic errors in the USNO~A2.0
$R$-band zero points.  For the remaining $\sim$10\% of sky patrol fields
without SDSS calibration data, we have offset the zero-points obtained from
USNO~A2.0 calibration by $0.22$ magnitudes.  For the remainder of this paper,
the magnitudes quoted are calibrated relative to the SDSS $r'$-band.

\begin{figure}
\rotatebox{90}{\scalebox{0.85}{\plotone{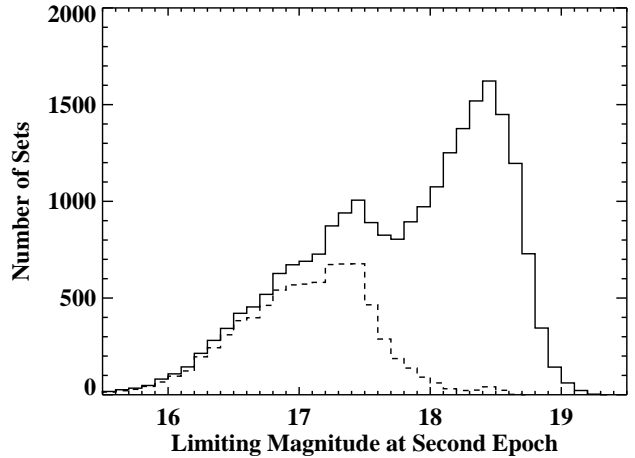}}}
\caption{\label{fig:wmlim}Histogram of number of \emph{sets} searched as a
  function of limiting magnitude ($m_{\mathrm{ROTSE(SDSS)}}$) at the second
  (return) epoch.  The limiting magnitude at the second epoch is the primary
  constraint for detecting quickly fading transients.  The solid histogram
  describes all \emph{sets} imaged, and the dashed histogram describes the
  contribution from 20~s exposures taken during bright lunar phases.}
\end{figure}

Through March 2005 we have searched over 23000 \emph{sets} of images for new
transients as described in \S~\ref{sec:intro}.  Figure~\ref{fig:wmlim} is a
histogram of the number of \emph{sets} searched as a function of limiting
magnitude at the second (return) epoch.  Since we demand that a new transient
be present in an entire \emph{set}, the limiting magnitude at the second epoch
is the primary constraint for detecting rapidly fading transients.  The solid
histogram describes all \emph{sets} imaged, and the dashed histogram describes
the contribution from 20~s exposures taken during bright lunar phases.

\section{Analysis}
\label{sec:analysis}

We have chosen to identify transients by the appearance of a new object
compared to a template list of ROTSE-III objects.  This strategy has distinct
advantages for our survey relative to image subtraction. As most of our fields
are relatively uncrowded, we only lose $\sim4\%$ of our solid angle to direct
starlight.  Due to bandwidth limitations at our remote observatory sites, we
need a strategy that is fast and has relatively few false positives.  Image
subtraction routines are very sensitive to variations in point spread function
(PSF) which are difficult to keep absolutely steady in an automated telescope
over a wide range of conditions.  Finally, as will be shown below, our solid
angle efficiency is over 80\% for detection of transients in most of our high
galactic latitude fields, comparable to the search strategy employed by the
DLS.~\citep{bwbcd04}

Our basic detection algorithm is quite simple.  We demand that a new object is
detected in at least four consecutive images, or an entire \emph{set}.  The
position resolution of each of the images must be less than 0.3 pixels
($1''$).  This removes images that are out of focus, and images with
anomalously large point spread functions common on windy nights. The new object
must be more than 5 pixels ($16\farcs2$) from the nearest ROTSE-III template
object.  We have no cuts on the shape of the light curve so as not to limit the
types of transient we can detect, although we do demand that the brightest
detection is brighter than $18^{th}$ magnitude.  For each candidate we examine
its PSF to ensure that it is comparable to the PSF of its neighbors.  We next
place an aperture at the same location in two of our best template images to
check if there is an object that had not been properly deblended by SExtractor.

After these cuts have been performed, typically $\ll$1 transient remains in each
\emph{set}.  Occasionally there are more detections, usually due to
instrumental effects such as artifacts near saturated stars, or due to
deblending problems caused by bad focus or wind-degraded images.  Therefore, if
more than 5 transient candidates remain in a \emph{set}, then it is rejected as
a bad \emph{set}.

Thumbnail images of the remaining transient candidates are then copied to a web
page at the University of Michigan for hand scanning.  These candidates are
usually faint stars that are just at our detection threshold, and are clearly
visible in the MAST Digitized Sky Survey.~\citep{lasker98} Occasionally there
is a minor planet near opposition that has a proper motion of $<1''$ during the
30 minute interval between exposure pairs. These minor planets are usually in
the Minor Planet Center MPChecker database, allowing easy exclusion.

In order to measure our overall efficiency, we must be able to parameterize our
coverage for each field.  We have used typical test fields to estimate the
detection efficiency as a function of distance to the nearest template star and
thus derive the effective solid angle covered for each of our sky patrol
fields.  We can also calculate the probability of detecting a transient
relative to the limiting magnitude of the field.

\begin{figure}
\rotatebox{90}{\scalebox{0.85}{\plotone{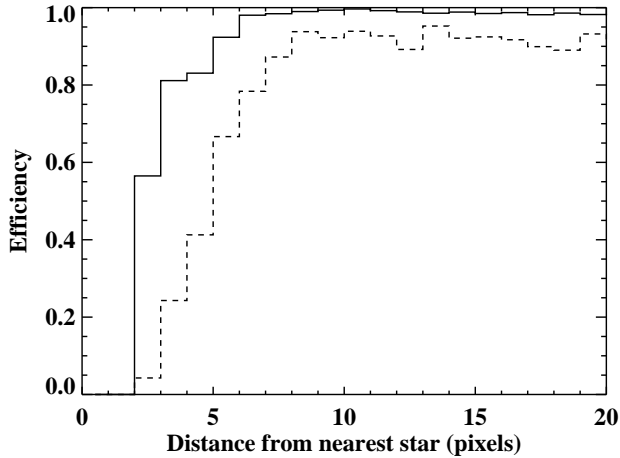}}}
\caption{\label{fig:effvsdist}Detection efficiency in a pair of images
  vs. distance from nearest template star.  The solid line describes simulated
  objects more than 0.5 magnitude brighter than the limiting magnitude; the
  dashed line describes simulated objects within 0.5 magnitude of the limiting
  magnitude, where our detection efficiency is reduced.}
\end{figure}

Figure~\ref{fig:effvsdist} shows the detection efficiency as a function of
distance to the nearest template star.  To obtain this plot we ran a Monte
Carlo simulation.  50000 objects were uniformly distributed in magnitude from
10.0 to 19.0 at random positions in a pair of sample images.  The objects were
generated with the IDL astronomy library function {\tt psf\_gaussian}, with the
median FWHM of a star in the field.  We then added the pixel counts at the
appropriate positions in the image.  These new images were reprocessed in the
standard analysis pipeline.  Only simulated objects that were detected in each
of the pair of images were considered.  For this plot, any object that is
within 2 pixels of a template star is vetoed; in our actual search we place the
cut at 5 pixels which incurs a minimal penalty in solid angle coverage while
greatly decreasing the number of false detections from deblending issues.  The
solid histogram describes simulated objects more than 0.5 magnitudes brighter
than the limiting magnitude.  The dashed histogram describes simulated objects
within 0.5 magnitudes of the limiting magnitude, where the detection efficiency
is reduced by $\sim$20\%.

\begin{figure}
\rotatebox{90}{\scalebox{0.85}{\plotone{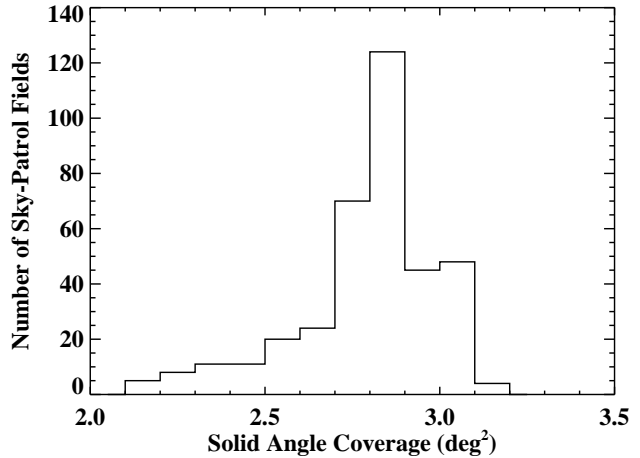}}}
\caption{\label{fig:areahist}Histogram of solid angle coverage for our sky
  patrol fields.  The maximum available solid angle in our field is
  $3.42\,\mathrm{deg}^2$, and the efficiency in most of the fields is $>$80\%.}
\end{figure}

Figure~\ref{fig:areahist} shows a histogram of the available solid angle
coverage for each of our sky patrol fields.  To determine these values, we
calculated the distance from each pixel to the nearest template star in each
field.  The solid angle lost in each field is primarily due to the 5 pixel
exclusion radius around each template star.  Additional pixels are masked out
in the wings of very bright stars and saturation bleed trails from bright
stars.  These typically cover $\sim$1\% of each field.  In most of our fields
the efficiency is greater than 80\%.  The low-coverage tail comprises dense
fields including those containing globular clusters.

\section{Results}

\subsection{Transient Detections}

Through March 2005, we have found three new cataclysmic variables, one flare
star, and one blazar.  This is a comprehensive list of all of our transient
detections that were not identified as asteroids. Each of these objects was
well below our detection limits in quiescence.  A brief description of these
objects follows.

\begin{itemize}
\item{CV ROTSE3 J151453.6+020934.2:} This CV was detected by ROTSE-IIIb on 28
  March, 2004, at 16$^{th}$ magnitude, over 3 magnitudes brighter than
  quiescence.  It remained around 17$^{th}$ magnitude for two weeks before
  fading below our detection threshold.  During quiescence we obtained $UBVI$
  measurements from the MDM Hiltner 2.4m telescope on Kitt Peak, Arizona.  Its
  colors were consistent with a dwarf nova at minimum light \citep{raagg04}.

\item{CV ROTSE3 J221519.8-003257.2:} This CV was detected by ROTSE-IIId on 8
  July, 2004, at 17.5 magnitude, around 3 magnitudes brighter than quiescence.
  The outburst lasted over two days before fading below our detection
  threshold.  As with the previous CV, its colors during quiescence were
  consistent with a dwarf nova at minimum light. \citep{raagg04}  This CV
  burst again on 4 October, 2004.  It was detected by ROTSE-IIId at 17.2
  magnitude, and was again discovered by our transient detection pipeline.

\item{Flare Star ROTSE3 J220806.9+023100.3:} This flare star was detected by
  ROTSE-IIIb on 12 November, 2004, at $16^{th}$ magnitude.  It faded by 0.8
  magnitudes in 30 minutes, mimicking our expected signature of an untriggered
  GRB afterglow.  However, the counterpart was clearly visible in 2MASS at
  $J\sim15$.  In addition, the $J-H$ and $H-K$ colors of 0.65 and 0.32
  respectively, suggest a very red object.  Finally, the USNO B-1.0 catalog
  measured proper motion of the quiescent counterpart of
  $32\,\mathrm{mas}/\mathrm{yr}$.  These observations are consistent with a
  flare from a nearby M-Dwarf type star.

\item{CV 2QZ J142701.6-012310:} This CV was detected by ROTSE-IIIc on 23
  January, 2005, at $15^{th}$ magnitude, around 5 magnitudes brighter than
  quiescence.  It remained bright around $16^{th}$ magnitude for about 7 days
  before fading below our detection threshold.  A spectrum previously had been
  obtained during quiescence by the 2dF redshift survey.  In addition, a
  spectrum of the object during outburst was obtained by the Hobby-Eberly
  Telescope (HET) at McDonald Observatory on 25 January, which showed a blue
  contiuum with no obvious emission or absorption features.~\citep{r05}
  Follow-up observations of the object during quiescence at the University of
  Cape Town revealed this to be a rare Am CVn type doubly-degenerate helium
  transferring binary.~\citep{wwr05}

\item{CV ROTSE3 J100932.2-020155:} This CV was detected by ROTSE-IIIc on 20
  February 2005, at $14^{th}$ magnitude, around 6 magnitudes brighter than
  quiescence. It faded slowly over the next 27 days before dropping below our
  detection threshold.  Two spectra had previously been obtained during
  quiescence by the 2dF redshift survey.  A spectrum from the HET on 21
  February displays prominent H-alpha emission expected from a CV, and the
  object is consistent with a SU UMa Dwarf Nova during a super
  outburst. \citep{rq05}

\end{itemize}

\subsection{Transient Detection Efficiency}

We calculate our detection efficiency using different methods for two different
types of transients.  The first method is suitable for slowly decaying
transients such as cataclysmic variables and other novae that rise rapidly and
are roughly constant over our observation interval.  These objects typically
stay brighter than our limiting magnitude for over 1 day.  Our effective time
coverage for these transients is therefore quite high.

In the second method, suitable for short duration (rapidly fading) transients,
we parameterize the transients as GRB afterglows with a simple decaying power
law.  This determines our sensitivity to rapidly varying transients but
describes a relatively small effective time coverage.  Specifically, the time
that the transient is brighter than our limiting magnitude must be $>30$
minutes.

To calculate our sensitivity for slowly decaying transients, we examine the
limiting magnitudes of each \emph{set} of images.  First, we account for the
solid angle covered, as plotted in Figure~\ref{fig:areahist}.  We then generate
5000 simulated transients uniformly distributed from 9.5 to $20^{th}$
magnitude.  If an object is $>$ 0.5 magnitudes brighter than the limiting
magnitude in an individual image it is considered a detection.  If it is $<$
0.5 magnitudes brighter than the limiting magnitude there is a 90\% chance that
it will be detected.  An object must be detected in all four of a \emph{set} of
images, and at least one detection must be brighter than $m_{ROTSE} = 18.0$.
This decreases our efficiency for faint CVs, but greatly decreases the false
detections as well as simplifying the interpretation of the data.

\begin{figure}
\rotatebox{90}{\scalebox{0.85}{\plotone{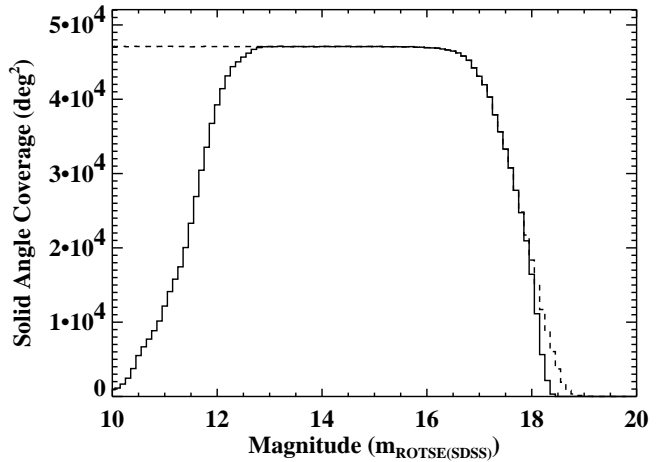}}}
\caption{\label{fig:staticcov}Histogram of total solid angle coverage for new
  slowly decaying sources.  The solid histogram describes objects that pass our
  cuts.  The dashed histogram describes objects that are detected but do not
  pass our cuts, either due to saturation at the bright end or due to falling
  below our magnitude threshold.}
\end{figure}

Figure~\ref{fig:staticcov} shows our total solid angle coverage for new slowly
decaying sources.  The detection efficiency before cuts is shown with the
dashed histogram.  After we apply saturation cuts and our magnitude cut, the
result is the solid histogram.  This plot was made using all of our \emph{sets}
with limiting magnitudes deeper than $\sim$17, with over
$47000\,\mathrm{deg}^2$ of coverage.  For transients that remain roughly
constant for over 0.5 days, this results in $\sim
52\,\mathrm{deg}^2\,\mathrm{yr}$ of coverage for CVs that peak between
$13^{th}$ and $16^{th}$ magnitude.

To calculate our sensitivity to short duration transients, we parameterize each
transient as a fading powerlaw, $f=f_{0}t^{-\alpha}$.  Each transient is
assigned a peak magnitude, $m_{60}$, at $t=t_0 + 60$s, and a decay constant
$\alpha$.  The effective coverage time for each \emph{set} is the time period
when a transient outburst would be detected in our search.  We have decided to
assume an effective coverage time of 30 m for each \emph{set} of four
observations, although our efficiency depends on peak magnitude and decay
constant for each transient.  The effective coverage time refers to the
\emph{preceding} 30 minute interval, as our search is insensitive to to
transients that appear between our two observation epochs. In order to achieve
$\sim100\%$ efficiency for a transient type in our observation window, it must
remain above our limiting magnitude for at least 1 hour.

As before with slowly decaying transients, we first account for the solid angle
covered in each field.  We ran a Monte Carlo simulation with 5000 objects per
\emph{set}.  The peak magnitude $m_{60}$ was uniformly distributed from 7.5 to
18.5, the decay constant $\alpha$ was uniformly distributed from 0.3 to 2.5,
and the burst time $t_0$ was set at a random time at most 30 minutes
\emph{prior} to the first image in the \emph{set}.  Essentially, we are
calculating the detection efficiency when assuming that our coverage time is 30
minutes for each \emph{set}.  As mentioned above, objects $<$ 0.5 magnitudes
brighter than the limiting magnitude have a 90\% chance of detection.

\begin{figure}
\rotatebox{90}{\scalebox{0.9}{\plotone{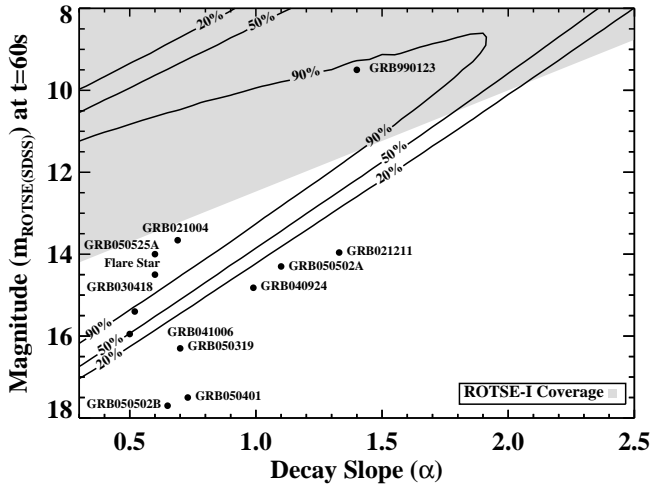}}}
\caption{\label{fig:transcov}Detection efficiency for fading transients,
  parameterized by their peak magnitude at $t=t_0+60$s, $m_{60}$, and their
  decay constant $\alpha$.  The efficiencies calculated assume 30 minutes of
  effective coverage for each \emph{set} for an integrated coverage of
  $1.74\,\mathrm{deg}^2\,\mathrm{yr}$.  All \emph{sets} with limiting
  magnitudes deeper than 17.5 were used for this plot.  The contours describe
  our 20\%, 50\%, and 90\% efficiency levels.  The contours roll over at the
  bright end where many transients would be saturated and rejected.  The gray
  area describes the coverage in the ROTSE-I transient search.  Overplotted are
  approximate peak magnitudes and decay constants (averaged over the first
  hour) for six GRB afterglows with early
  detections.~\citep{abbbb99,fyktk03,fox04,lfcj03,rspaa04,rsq05,rys05c,rys05,yssa05,fbcsc05}}
\end{figure}

Figure~\ref{fig:transcov} shows our detection efficiency for all the
\emph{sets} in the search with limiting magnitudes at the second epoch deeper
than 17.5, assuming 30 minutes of coverage. The integrated coverage is
$1.74\,\mathrm{deg}^2\,\mathrm{yr}$ for these \emph{sets}.  The 20\%, 50\%, and
90\% contours are shown.  The contour lines roll over at the bright end where
our saturation cuts take effect -- these extraordinarily bright transients
would be vetoed in our pipeline.  However, the ROTSE-I transient search covered
much of this parameter space (shaded in gray) without finding any candidate
afterglows with approximately twice as much coverage. \citep{kabbc02}
Overplotted are approximate peak magnitudes and decay slopes, averaged over the
first hour after the burst time, for ten of the twelve GRB afterglows that have
been detected in the first hour after the burst.  The remaining two bursts are
too faint to be placed on this plot.  We are sensitive to $\sim40\%$ of these
afterglows, which have been detected for $\sim50\%$ of all promptly localized
GRBs.  Also plotted is an inferred peak magnitude and decay slope for the flare
star ROTSE3 J220806.9+023100.3, which is contained in the locus of afterglow
points.

\section{Discussion}

To date, this is the deepest wide-field search for untriggered and orphan GRBs.
With a total coverage of $1.74\,\mathrm{deg}^2\,\mathrm{yr}$, we have not found
any candidate afterglows or other unknown transients.  The primary reason we
have been able to positively identify each of our transient candidates is the
5-color SDSS data of the fields in our search.  Other transient searches such
as the DLS that can image much deeper do not have this advantage.

We have also determined that the backgrounds for a search of this type are
purely astrophysical.  We have not had any difficulty from satellite glints or
any mysterious terrestrial flashes.  Our primary background consists of
asteroids with small proper motion, but the MPChecker database combined with
follow-up observations can eliminate these.  Outside the solar system, but
within the galaxy, we see a small but measurable rate of CVs and flare stars,
as well as one extragalactic blazar.  The flare star ROTSE3 J220806.9+023100.3
is the only object that we have found that had a light curve that mimicked our
expected signature of an untriggered GRB afterglow.

We can place an upper limit on the rate of fading optical transients with
quiescent counterparts dimmer than $\sim20^{th}$ magnitude, as described by our
90\% coverage region shown in Figure~\ref{fig:transcov}.  To calculate an upper
limit on the rate of transients in this region, we follow the method of
\citet{bwbcd04}.  The observed rate is
\begin{equation}
\eta =
\frac{N}{{\langle}{\cal E}{\rangle}E}\,\mathrm{events}\,\mathrm{deg}^{-2}\,\mathrm{yr}^{-1}\label{eqn:rate},
\end{equation}
where $N$ is the number of events, $E$ is the exposure, and ${\langle}{\cal
E}{\rangle}$ is the efficiency.  With the observed number of transients
$N_{\mathrm{obs}} = 0$, Poisson statistics place an upper limit of
$N_{\mathrm{max}} < 3.0$ with 95\% confidence.  Therefore, we can place a 95\%
confidence upper limit of $\eta_{\mathrm{max}} <
1.9\,\mathrm{deg}^{-2}\,\mathrm{yr}^{-1}$ in our coverage region.

The launch of Swift will allow us to probe the early afterglow phase of GRBs
far more systematically than has been achieved to date.  We have learned some
details from the variety of early afterglows detected so far.  It is clear that
very bright prompt flashes like that from GRB~990123~\citep{abbbb99} are not
the norm.  It also appears that extrapolating late time afterglows to the early
time generally over-predicts the brightness, as with GRB~030418 and
GRB~030723.~\citep{rspaa04}  These data certainly make our search more
difficult, as only $\sim20\%$ of GRBs have afterglows that fall within our
sensitivity region.

If we assume that the $\gamma$-ray emission of GRBs is confined within a double
jet with a cone half-angle of $\theta_{\mathrm{max}}$, while the optical
emission is isotropic, then a given GRB is visible in a small fraction of the
sky approximated by $\theta^2/2$.  Therefore, the true rate of GRB events
within the observable universe must be
$\sim1500/\theta_{\mathrm{max}}^2\,\mathrm{events}\,\mathrm{yr}^{-1}$.  The
95\% confidence limit of $<78000\,\mathrm{events}\,\mathrm{yr}^{-1}$ is for a
region that is sensitive to $\sim20\%$ of GRBs, and therefore our assumption of
isotropic optical emission is tenable as long as
$\theta_{\mathrm{max}}>3.6^{\circ}$.  As this is an approximate estimate of the
GRB jet angle, the present limits cannot set stringent bounds on the properties
of these objects.  However, if programs such as ours continue to reduce the
upper bounds for the orphan afterglow rate, the isotropic emission hypothesis
will become incompatible with what we know about the structure of GRB jets.
Although this is not an enormous surprise, it does represent a sanity check of
the accepted model of GRBs by completely independent reasoning.

Our search has also detected several previously unknown high galactic latitude
cataclysmic variables with dim quiescent counterparts.  We have detected 3 new
CVs with $\sim52\,\mathrm{deg}^2\,\mathrm{yr}$ of coverage for slow decay
transients that peak between $13^{th}$ and $16^{th}$ magnitude.  This implies a
rate of $0.06\,\mathrm{deg}^{-2}\,\mathrm{yr}^{-1}$.  Assuming Poisson
statistics, the 95\% confidence upper limit on the rate is
$0.17\,\mathrm{deg}^{-2}\,\mathrm{yr}^{-1}$.  We have found 1 new CV with
$\sim25\,\mathrm{deg}^2\,\mathrm{yr}$ of coverage for CVs that peak between
$16^{th}$ to $18^{th}$ magnitude.  This implies a rate of $0.04\,
\mathrm{deg}^{-2}\,\mathrm{yr}^{-1}$.  Assuming Poisson statistics, the 95\%
confiedence upper limit on the rate is
$0.21\,\mathrm{deg}^{-2}\,\mathrm{yr}^{-1}$.  If we extrapolate the rate over
the whole sky, this would produce $\sim2000$ high galactic latitude CV
outbursts every year.  In the ``Living Edition'' catalog of CVs there are only
$\sim 130$ CVs in our sensitivity range as of March 2005.~\citep{dwsrk01} None
of the CVs detected by ROTSE-III had been known previously.  Our search results
imply that many more CVs remain to be discovered.

The ROTSE-III transient search is an ongoing project.  We expect to continue to
patrol the sky over the lifetime of the Swift instrument (at least 2 years), as
we wait for GRB triggers.  We will thus gain another factor of 2-3 in coverage,
and will achieve a modest improvement towards the goal of
$\sim110\,\mathrm{deg}^2\,\mathrm{yr}$, the threshold for having a high
probability of finding an untriggered afterglow.  In this paper, we have
demonstrated that such a search is feasible, as the background rate of unknown
transients is very low.

\acknowledgements

This work has been supported by NASA grants NNG-04WC41G and F006794, NSF grants
AST-0119685 and 0105221, the Australian Research Council, the University of New
South Wales, and the University of Michigan.  Work performed at LANL is
supported through internal LDRD funding.  Special thanks to the
observatory staff at the ROTSE sites, including David Doss, Toni Hanke, and
Tuncay \"{O}z{\i}\c{s}{\i}k.

\end{document}